\documentclass[journal]{IEEEtran}
\usepackage{graphicx}
\usepackage{epstopdf}
\usepackage{amsfonts,amsmath,amssymb,amsthm,caption}
\usepackage[ruled,vlined,commentsnumbered]{algorithm2e}
\usepackage{algpseudocode}
\usepackage{amsbsy}
\usepackage{graphicx,multirow,bm}
\usepackage{color}
\usepackage{amsmath}
\usepackage{cite}
\usepackage{pstricks}
\newtheorem{example}{Example}
\newtheorem{theorem}{Theorem}
\newtheorem{lemma}{Lemma}

\begin{document}
\title{Multiaccess Coded Caching with Private Demands}

\author{Dequan Liang,
        Kai Wan,~\IEEEmembership{Member,~IEEE,}
        Minquan Cheng,
        Giuseppe Caire,~\IEEEmembership{Fellow,~IEEE}
\thanks{D. Liang, M. Cheng are with Guangxi Key Lab of Multi-source Information Mining $\&$ Security, Guangxi Normal University,
Guilin 541004, China  (e-mail: dequan.liang@hotmail.com, chengqinshi@hotmail.com).}
\thanks{K. Wan and G. Caire are with the Electrical Engineering and Computer Science Department, Technische Universit\"{a}t Berlin,
10587 Berlin, Germany (e-mail: kai.wan@tu-berlin.de, caire@tu-berlin.de).  The work of K.~Wan and G.~Caire was partially funded by the European Research Council under the ERC Advanced Grant N. 789190, CARENET.}}


\maketitle

\begin{abstract}
  Hachem et al. formulated a multiaccess coded caching model which consists of a central server connected to $K$ users via an error-free shared link, and $K$ cache-nodes. Each cache-node is equipped with a local cache and each user can access   $L$ neighbouring cache-nodes in a cyclic wrap-around fashion. In this paper, we take the privacy of the users' demands into consideration, i.e., each user, while retrieving its own demanded file, cannot obtain any information on the demands of the other users.
 By storing some private keys at the cache-nodes, we develop a novel transformation approach to turn  any  non-private coded caching scheme (satisfying some constraints) into a private one.
%{\red COMMENTS OF KAI: USE ONE SENTENCE TO SUMMARIZE THE STRATEGY OF THE SCHEME.}
 %By adding random vectors, we show that each multiaccess coded caching scheme satisfying symmetric across files can be used to generate a multiaccess coded caching scheme with private demands.
\end{abstract}



% \fi
\IEEEpeerreviewmaketitle

\section{Introduction}
\label{sec-introduction}
Recently, as the wireless data traffic is increasing at an incredible rate dominated by the video streaming,   wireless networks are subject to a tremendous capacity stress. Consequently, communication systems are likely to be congested during the peak traffic times. Reducing this congestion is very desirable  and represents a hot topic in industrial applications and academic research. Coded caching, originally proposed in~\cite{MAN}, is an efficient solution to turn storage into bandwidth resource, in order to reduce such congestion for ``cachable'' content, as for example in the case of video on demand.
%\IEEEPARstart{W}{ith} a large number of wireless communication devices connecting to the Internet, Internet data traffic is increasing rapidly. The current wireless communication networks are increasingly under pressure to effectively transmit data traffic. Caching is a promising technology that can effectively alleviate the pressure of data transmission. In recent years, the pioneering coded caching, which was originally proposed by Maddah-Ali and Niesen, has received significant interest.
 In the shared-link coded caching model~\cite{MAN} (also called MN model),   a server with access to a library of $N$   files,  is connected to $K$ users via an error-free shared link. Each user is equipped with a local cache of $M$ files. A coded caching scheme includes two phases: {\it placement phase} and {\it delivery phase}. In the placement phase, without knowledge of future demands, the server fills part of the popular files into the user's cache. In the delivery phase, each user demands a file from server. After receiving the users' demands, the server sends multicast  messages such that each user's demand can be satisfied. The goal is to minimize the amount of data transmission (referred to as {\it load}) in the delivery phase. Inspired by the seminal work in~\cite{MAN}, the caching problem for networks with various topologies has been studied, e.g., Device-to-Device networks in \cite{JCM}, combination networks in \cite{JAJC,WJPD} and Hierarchical networks in \cite{NUMS}, etc.

\begin{figure}
\centerline{\includegraphics[scale=0.5]{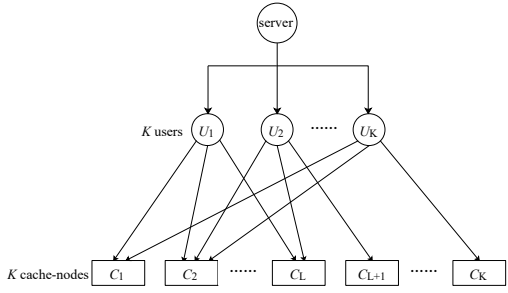}}
\caption{\small The $(K,L,M,N)$ multiaccess  coded caching model.}
\label{multiaccess-system}
\end{figure}

In the MN coded caching scheme, each user's demand is globally known, so %users'   preferences will be exposed to other users, and
the privacy of the users' demands can not be guaranteed~\cite{Engle2017privatecaching}. The information theoretic coded caching model with private demands  was formulated in  \cite{WG} for the shared-link model.  The authors in \cite{WG} introduced virtual users on the shared-link model and used the delivery strategy of MN scheme to satisfy the demands of all users including real users and virtual users, such that each real user can not distinguish the demands of other real users. The virtual-user scheme was proven to be order optimal within a constant factor except for the regime where $N>K$ and $M<N/K$.  %\cite{WHMDG} formulated a Device-to-Device (D2D) private caching model with a trusted server, which receives the user's request and then sends a query to  each user for the further transmission by this user. Similarly, in order to preserve the privacy of the users' demands, the strategy of adding virtual users was used.
Following the new formulation in~\cite{WG}, variants of the private caching scheme were proposed in~\cite{VPA,SJB} to improve the virtual user scheme in~\cite{WG}, in terms of either the transmission load or the sub-packetization level.
 Very recently, a private coded caching scheme was proposed in  \cite{YT} which is based on the cache-aided linear function retrieval scheme in~\cite{arxivfunctionretrieval}.
  The key idea of the scheme in \cite{YT} is that, in addition to cached contents by the MN placement strategy, users  privately cache some linear combinations of uncached subfiles in the MN placement  which are regarded as keys. In such way, the effective demand of  each user in the delivery phase become the sum of these linear combinations and the   subfiles  of its desired file, such that the real demand is concealed.  The scheme in \cite{YT} has a similar load as the virtual-user scheme in \cite{WG}, but with a much lower sub-packetization level.
%{\red COMMENTS OF KAI: YOU NEED A MORE FORMAL AND DETAILED REVIEW ON THE EXISTING PRIVATE CODED CACHING SCHEMES; TRY TO USE ONE OR TWO SENTENCES FOR EACH ONE TO SUMMARIZE ITS MAIN IDEA AND  ITS ADVANTAGE (OR ORDER OPTIMALITY GUARANTEE).}
     %  privacy issue in caching schemes have been widely concerned. \cite{WG} formulated a caching system with private demands , which aims to preserve the privacy of the demand of each user from other users.
   %  Then the problem of coded caching with private demands for Devide-to-Device (D2D) networks was considered in \cite{WHMDG}, that is, each user can not get any information about the demands of other use.
     %Reference \cite{YT} is the latest work that studied the similar privacy problem in MN caching model. The private caching scheme proposed in \cite{YT} was proved to be order optimal and has lower sudpacketization comparing with the strategy of virtual users.
\paragraph*{Multiaccess caching system}
The multiaccess   caching model (illustrated in Fig.~\ref{multiaccess-system}),  originally proposed in~\cite{JHNS},  is motivated by the concept of ``edge caching'', where caching nodes (separated from the users) can be accessed simultaneously by multiple users.
 Different from the MN caching model, in the multiaccess caching model, there are $K$ cache-nodes each of which can store $M$ files, and $K$ cache-less users each of which can access $L$ cache-nodes in a cyclic wrap-around fashion, while also receiving the broadcast transmission directly from the server. As assumed in \cite{JHNS}, the cache-nodes in this model are accessed at no load cost; that is, we aim to minimize  the broadcasted load from the server for any given $M$. The authors in \cite{JHNS} proposed a scheme, which achieves the load $\frac{K(1-LM/N)}{ KM/N+1}$ when $L$ divides $K$. In \cite{CLWZG} the authors proposed a transformation approach to extend some schemes for the MN model to the multiaccess caching system. As an application, two new multiaccess coded caching schemes were obtained based on the MN scheme and partition scheme in \cite{YCTC}.
It is worth noting that the first scheme in \cite{CLWZG} achieves the transmission load $\frac{K(1- LM/N)}{ KM/N+1}$ for any positive integers $L$ and $K$. In~\cite{RK,SR,SPE,MARB}, other constructions of multiaccess coded caching schemes were proposed based on uncoded cache placement (i.e., each cache-node only cache subset of library bits) and linear delivery coding. Recently, the authors in \cite{KKKBS} studied secure multiaccess caching problem which aims to protect the security of the library content from some wiretapper.

\paragraph*{Contribution and paper organization}
Different from the secure multiaccess caching problem in \cite{KKKBS}, in this paper we formulate the multiaccess caching model with private demands, which aims to preserve the privacy of each user's demand from other users.
 Compared to the shared-link caching model with private demands in \cite{WG}, the main challenge to preserve the demand privacy in the multiaccess caching model is that the users share the cached contents at the cache-nodes while in the shared-link model, the existing private caching schemes are designed based on the fact that the composition of the   each user's cache is unknown to the other users.

Besides the novel problem formulation, our main contribution is to propose an approach to add the privacy guarantee for
   any non-private multiaccess coded caching scheme  satisfying three requirements,  such as the non-private schemes in~\cite{CLWZG,RK}. Based on the feature that each user is connected to $L$ neighbouring cache-nodes in a cyclic wrap-around fashion, we   use second-layer  keys to protect the first-layer keys (i.e., the private linear combinations as in \cite{YT}), so that even if users share some cached contents, each user can privately obtain these first-layer keys.
  %  {\red THIS IS THE MAIN NOVELTY OF THE SCHEME in \cite{YT}, NOT OURS! TRY TO SUMMARIZE THE MAIN NOVEL OF OUR SCHEME COMPARED TO THE SCHEME IN \cite{YT} AND EXPLAIN HOW TO OVERCOME THE CHALLENGE WHICH I WROTE IN THE LAST PARAGRAPH!}}

 \iffalse
\paragraph*{Paper organization}
%The rest of this paper is organized as follows.
Section \ref{sec-system-model} formulates the considered problem.  Section \ref{sec-result} presents the main results of the paper, while  the proposed scheme is described in section \ref{sec-proposed}.   Section \ref{sec-conclusion} concludes the paper.
 \fi
%Mod $(b,a)$ represents the modulo operation on positive integer $b$ with positive integer divisor $a$.

\paragraph*{Notations}
In this paper,  we use the following notations unless otherwise stated.
Bold capital letter, bold lower case letter and curlicue  font  will be used to denote matrix, vector and set, respectively. $[a:b]:=\left\{ a,a+1,\ldots,b\right\}$ and $[n] := [1:n]$; $|\cdot|$ represents the cardinality of a set or the length of a vector. Mod $(b,a)$ represents the modulo operation on positive integer $b$ with positive integer divisor $a$.  In this paper we let $\text{Mod}(b,a)\in \{1,\ldots,a \}$ (i.e., we let $ \text{Mod}(b,a)=a$ if $a$ divides $b$).
%; $\mathbb{F}_{q}$ represents a finite field with order $q$.
%In this paper we let $\text{Mod}(b,a)\in \{1,\ldots,a \}$ (i.e., we let $ \text{Mod}(b,a)=a$ if $a$ divides $b$).

\section{System Model}
\label{sec-system-model}%
%\subsection{System Model}
%\label{sub:system model}
A   $(K,L,M,N)$ multiaccess caching problem with private demands  has a single server containing $N$ independent files $W_{1}, W_{2}, \ldots, W_{N}$, $K$ cache-nodes $ C_1, C_2,\ldots,C_{K}$, and $K$ users $ U_{1}, U_2, \ldots, U_{K}$. The server is connected to $K$ users via an error-free shared link. We assume that each file is composed of $B$  i.i.d. bits.

%A $(K,L,M,N)$ multiaccess caching system with private demands can be introduced as follows.\\
\indent {\bf Placement phase.} The server fills  the cache of each cache-node without   knowledge of the users' future demands. We denote the contents cached at the cache-node $C_{k}$ by $\mathcal{Z}_{C_{k}}$.
In order to preserve the privacy of each user's demand, the server   generates $K$  random variables, $V_1 $ over  $\mathcal{V}_1$, $V_2 $ over  $\mathcal{V}_2$,  $\ldots$, $V_K$ over  $\mathcal{V}_K$, which are independent of the library files. %Then server maps the files $W_{1}, W_{2}, \ldots, W_{N}$ and the key sets into the cache memory of each cache-node $C_{k}$,
Hence,
$$
\mathcal{Z}_{C_{k}} = \phi_{C_{k}}(W_{1}, W_{2}, \ldots, W_{N}, V_k)
$$
where $\phi_{C_{k}}:[\mathbb{F}_{2}]^{N B} \times  \mathcal{V}_k    \rightarrow [\mathbb{F}_{2}]^{M B}$.
Each user $U_{k}$ where $k \in [K]$  can retrieve the contents of $L \leq K$ neighbouring cache-nodes in a cyclic wrap-around fashion, i.e., the cache-nodes in $\{C_{\text{Mod}(k,K)}, C_{\text{Mod}(k+1,K)}, \ldots, C_{ \text{Mod}(k+L-1,K)} \}$. Define
$$
\mathcal{Z}_{U_k}:= \cup_{i   \in \{ \text{Mod}(k,K), \text{Mod}(k+1,K), \ldots, \text{Mod}(k+L-1,K) \}} \mathcal{Z}_{C_{i}}.
$$
 as  the retrievable contents by user $U_k$.
We assume that each user can retrieve the cached contents from its connected cache-nodes without any cost.
\\
\indent {\bf Delivery phase.} Assume that the demand vector is $\mathbf{d}=(d_{1},d_{2},\ldots,d_{K})$. User $U_{k}$, $k \in [K]$, demands the file $W_{d_{k}}$ from server.
Notice that the users' demands are    independent of the library contents.
\iffalse
\begin{align}
 H\big(W_{1},\ldots,W_{N},  \mathbf{d} \big)& = H(W_1)+\cdots+H(W_N) +H(\mathbf{d}).
\end{align}
\fi
With the knowledge of ${\bf d}$, the server broadcasts the messages
$$
X = \psi(\mathbf{d}, W_{1},  \ldots, W_{N}, V_1,   \ldots,V_K),
$$
where $\psi:[N]^{K} \times [\mathbb{F}_{2}]^{NB} \times \mathcal{V}_1\times \cdots \times \mathcal{V}_K  \rightarrow [\mathbb{F}_{2}]^{RB}$ and $R$ is called as the transmitted load or load. With the help of the received messages and the retrievable cached contents, each user $U_{k}$ where $k \in [K]$  must decode its demanded file $W_{d_{k}}$. Hence for any user $U_{k}$ it must hold that%\footnote{\label{foot:metadata}As in~\cite{}, for the sake of successful decoding, we assume that the metadata/composition of   $\mathcal{Z}_{C_{k}}$ (i.e., from which code on which bits,    $\mathcal{Z}_{C_{k}}$  are generated) is also provided  in $\mathcal{Z}_{C_{k}}$, for each $k \in [K]$. Similarly, the metadata of $X$ is also provided in $X$. Since the length of the metadata is negligible compared to the information packets, we do not specify them in the rest of the paper.}
\begin{eqnarray}
\label{eq-user-demand}
H(W_{d_{k}} | X, d_{k}, \mathcal{Z}_{U_k}) = 0.
\end{eqnarray}
In addition, we impose a privacy constraint on the multiaccess caching model, i.e., each user  can not get any information about other users' demands from $X$, $\mathcal{Z}_{U_k}$ and $d_k$; that is,
\begin{eqnarray}\label{eq-privacy}
I({\bf d}_{-k}; X ,\mathcal{Z}_{U_k}| d_{k}) = 0, \ \forall k\in [K],
\end{eqnarray}where ${\bf d}_{-k}=(d_{1},\ldots,d_{k-1},d_{k+1},\ldots, d_{K})$ denotes the demands of all the users except the demand of the user $U_{k}$.

%Assume that the communication bottleneck is on the shared link from the server to the users.
By the constraint of privacy, we can see that the loads for different demand vectors should be the same;  otherwise, the transmitted load which can be counted by each user will reveal information about the other users' demands.  The objective of this paper is to minimize the    load  $R$  by designing  a  placement and delivery strategy that satisfies each user's demand in \eqref{eq-user-demand} and the privacy constraint in \eqref{eq-privacy}.

\section{Main Result}
\label{sec-result}
We will propose a general approach in Section~\ref{sec-proposed} to add the privacy guarantee for any non-private multiaccess coded caching scheme satisfying three requirements.
\begin{theorem}
\label{th-one}
From any $(K,L,M',N)$ non-private  multiaccess coded caching  scheme  with $L< K/2 +1$ satisfying the following   requirements,
\begin{itemize}
\item {\bf Requirement~1:} Any $L$ neighbouring cache-nodes do not cache any common bits;
\item {\bf Requirement~2:} The placement phase is {\em identically uncoded}, i.e., if some cache-node stores the $i$-th bit of some file, then it caches the $i$-th bit of all files.
\item {\bf Requirement~3:} In the delivery phase, it treats the demand of each user as an independent file.
\end{itemize}
we obtain a $(K,L,M,N)$ private multiaccess coded caching scheme  where $M=M'+2(1-\frac{LM'}{N})$ with the same load and sub-packetization level as the original non-private scheme.
\hfill $\square$
\end{theorem}

The proof of Theorem \ref{th-one} and some detailed examples are provided in Section \ref{sec-proposed}. Using the same placement and delivery strategies in Section \ref{sec-proposed} and   Shamir's secret sharing in \cite{Sharmir},
 we can prove a demand-private scheme also in the regime $L \geq K/2+1 $. In particular,
for any non-private scheme $(K,L,M',N)$ satisfying the requirements in Theorem \ref{th-one}, we can construct a demand-private scheme with the same load and memory
$M = M' + \omega (1 - LM'/N)$ where $\omega$ is an integer $\geq  2$ which will be clarified   in Section~\ref{sec:other region}.

From Theorem~\ref{th-one} it follows that introducing the additional  feature of demands' privacy to a non-private multiaccess coding scheme satisfying Requirements 1, 2, and 3 incurs a cost of $M-M'= 2(1-\frac{LM'}{N}) \leq 2$ in cache size.
In other words, the additional  memory size is upper bounded by an additive constant factor of $2$. 

By comparing the requirements in Theorem~\ref{th-one} and the existing non-private multiaccess coded caching schemes   in~\cite{JHNS,CLWZG,RK,SR,MARB,SPE},
it can be seen that all of those schemes satisfy {\bf Requirements~1, 2, 3}   in Theorem~\ref{th-one}. For simplicity, here we only take the following scheme in \cite{CLWZG} as an example. 
\begin{lemma}[\cite{CLWZG}]\rm
\label{lemma-1}
For the $(K,L,M',N)$  non-private multiaccess coded caching problem,
the lower convex envelope of the following memory-load tradeoff   points are achievable,
\begin{align*}
(M', R_1)= \left(\frac{N t}{K}, \frac{K-tL}{t+1}\right), \ \forall t \in \left[0:  \left\lfloor
\frac{K}{L} \right\rfloor \right] , %\label{eq:load R1}
\end{align*}
and $(M',R_1)=\left( \frac{N}{L}, 0\right)$.
\end{lemma} 
Based on the scheme in Lemma \ref{lemma-1} and by Theorem~\ref{th-one}, we can obtain the following theorem. 
 \begin{theorem}\rm
\label{theorem-2}
For the $(K,L,M,N)$ multiaccess  coded caching problem with private demands,
the lower convex envelope of the following memory-load tradeoff points are achievable,
\begin{align*}
&\left(M, R_{1}\right)  =  \left(\frac{(N-2L) t}{K}+2, \frac{K-tL}{t+1} \right),\ \forall t  \in  \left[ \left\lfloor
\frac{K}{L} \right\rfloor  \right] ,  
\end{align*} if $L< K/2 +1$, and $(M,R_1)=\left( \frac{N}{L}, 0\right)$.
\hfill $\square$
\end{theorem}

\iffalse
%{\red YOU HAVE THE SPACE TO ADD SOME NUMERICAL EVALUATIONS. CHOOSE ONE CASE, PLOT THE EXISTING NON-PRIVATE MULTIACCESS CACHING SCHEMES AND OUR PRIVATE SCHEME.}
 In Fig.~\ref{Figure-compare}, we provide a numerical example for the   multiaccess coded caching problem with private demands with $N=40$, $K=20$, and $L=3$, where we compare non-private multiaccess coded caching scheme  in Lemma~\ref{lemma-1} and the proposed private multiaccess coded caching scheme  in Theorem~\ref{theorem-2}. It can be seen that the to guarantee the privacy with the same load as the original non-private   scheme in Lemma~\ref{lemma-1}, the needed additionally memory size of proposed private   scheme is less than $2$.

\begin{figure}%[ht]
%\vspace{-2mm}
\centerline{\includegraphics[scale=0.3]{Figure_K20_L3}}
\caption{\small The private multiaccess  coded caching system with $N=40$, $K=20$, and $L=3$.}
\label{Figure-compare}
%\vspace{-5mm}
\end{figure}
\fi

\section{Proof of Theorem~\ref{th-one}}
\label{sec-proposed}
Assuming that there exists a $(K,L,M',N)$ non-private multiaccess coded caching scheme, say Scheme 1, which satisfies {\bf Requirement~1} and {\bf Requirement~2}.
\iffalse
 For any $n \in [N]$, define that
  \begin{align}\label{eq-file-packet}
  W_{n}=\{W_{n,j}| \ \forall j \in [B]\}.
  \end{align}
  \fi
Let us then introduce the corresponding  $(K,L,M=M'+2(1-\frac{LM'}{N}),N)$ private multiaccess coded caching scheme.
\subsubsection{Placement Phase}
\label{subsect-placement}
\iffalse
  The files are split in the same way as Scheme 1. As the placement of Scheme 1 is identically uncoded,   we can divide each file $W_n$ where $n\in [N]$ into subfiles,
$
W_{n}=\{ W_{n,\mathcal{S}} | \mathcal{S} \subseteq [K]\},
$
  where $W_{n,\mathcal{S}}$ represents the set of bits in $W_n$  which are uniquely cached by cache-node in $\{C_i| i\in \mathcal{S} \}$.
  \begin{align*}
  W_{n}=\{W_{n,j}| \ \forall j \in [B]\}.
  \end{align*}.
  \fi
  Each cache-node caches two parts of cached contents. The first part  is the same as  the contents cached at cache-nodes in the placement phase of  Scheme 1.
As the placement of Scheme 1 is identically uncoded,   we can divide each file $W_n$ where $n\in [N]$ into subfiles,
$
W_{n}=\{ W_{n,\mathcal{Q}} | \mathcal{Q} \subseteq [K]\},
$
  where $W_{n,\mathcal{Q}}$ represents the set of bits in $W_n$  which are uniquely cached by cache-nodes in $\{C_i| i\in \mathcal{Q} \}$.
Since the placement phase is identically uncoded,  we define $x_{\mathcal{Q}}=|W_{1,\mathcal{Q}}|=\cdots=|W_{N,\mathcal{Q}}|$.

 For each $k\in [K]$, we define
$
   \mathcal{T}_{k} = \Big\{\mathcal{Q} \subseteq [K] | \mathcal{Q} \cap \{ \text{Mod}(k,K),   \ldots, \text{Mod}(k+L-1,K)\}= \emptyset \Big\}.
$
Hence, each subfile $W_{n,\mathcal{Q}}$ where $\mathcal{Q} \in    \mathcal{T}_{k} $ cannot be retrieved by user $U_k$ from its connected cache-node.
%{\red KAI STOPPED HERE. MODIFY THE FOLLOWING DESCRIPTION USING THE CONCEPTS OF SUBFILE AND THE NEW $T_k$.}

From {\bf Requirement~1} and {\bf Requirement~2}, we have for each user $k \in [K]$, the number of bits of each file which user  $k$ cannot retrieve from its connected cache-nodes is
$$
\sum_{\mathcal{Q} \in    \mathcal{T}_{k}} x_{\mathcal{Q}} = (1-L M'/N) B.$$

%$|\mathcal{T}_{1}|=|\mathcal{T}_{2}|=\ldots=|\mathcal{T}_{K}|$.

In  the second part, by the following two steps, each cache-node additionally stores  some linear combinations of subfiles as the `keys' which can be retrieved by the connected users.
      \begin{itemize}
      \item {\it Step 1.}  The server   randomly generates $K$ vectors where $\mathbf{p}_{k}=(p_{k,1},p_{k,2},\ldots,p_{k,N}) \in \mathbb{F}_{2}^N$, $k\in [K]$, which  are independent of the library files.
       Then for any $\mathcal{Q} \in \mathcal{T}_{k}$, the server generates a linear combination
      \begin{align}\label{eq-linear}
      S_{k,\mathcal{Q}}=   \underset{n\in [N]}{\oplus }   p_{k, n} W_{n,\mathcal{Q}}.
      \end{align}
      \item  {\it Step 2.}
For each $k\in [K]$ and each $\mathcal{Q} \in \mathcal{T}_{k}$, the server generates a random variable $A_{k,\mathcal{Q}}$, which is uniformly i.i.d. over   $\mathbb{F}^{x_{\mathcal{Q}}}_{2}$ and is independent of the library files.
      Then for each $k\in [K]$, the server generates the key set
\begin{eqnarray*}
 \mathcal{P}_k&=& \{S_{k,\mathcal{Q}} \oplus A_{k,\mathcal{Q}} |\forall \mathcal{Q} \in \mathcal{T}_{k}\} \bigcup  \\
       &&   \left\{  A_{\text{mod}(k-L+1,K),\mathcal{Q}}\ |\forall \mathcal{Q} \in \mathcal{T}_{\text{mod}(k-L+1,K) }\right\}
\end{eqnarray*}  and places it into the cache of cache-node $C_{k}$.
\end{itemize}

Each cache-node  totally caches $M'B+2(1-L M'/N) B =M B$ bits, satisfying the memory size constraint.

For each $k\in [K]$, user $U_{k}$ can retrieve $S_{k,\mathcal{Q}} \oplus A_{k,\mathcal{Q}}$ where $\mathcal{Q} \in \mathcal{T}_{k}$ from $C_{k}$. Similarly,   user $U_{k}$ can also obtain $ A_{k,\mathcal{Q}}$ where $\mathcal{Q} \in \mathcal{T}_{k}$ from cache-node $C_{\text{mod}(k-L+1,K)}$.\footnote{\label{foot:two different}Since $L \leq K$, we have $L-1<K$. Hence, cache-nodes $C_{k}$ and $C_{\text{mod}(k-L+1,K)}$ are two different cache-nodes.} Hence it can successfully recover  $S_{k,\mathcal{Q}}$ where $\mathcal{Q} \in \mathcal{T}_{k}$. Let us define that $\mathcal{S}_{k}:=\{S_{k,\mathcal{Q}} | \forall \mathcal{Q} \in \mathcal{T}_{k}\}$. For each $j\in [K] \setminus \{k\}$ and each $\mathcal{Q} \in \mathcal{T}_{j}$, $S_{j,\mathcal{Q}}\oplus A_{j,\mathcal{Q}}$ is only cached by cache-node $C_{j}$, while $A_{j,\mathcal{Q}}$ is only cached by cache-nodes $C_{\text{mod}(j+ L-1)}$. So except user $U_{j}$, none of the other users is simultaneously connected to cache-nodes $C_{j}$ and  $C_{\text{mod}(j+ L-1)}$ if $L< K/2 +1$. Hence, we have the following key point for our scheme:\\[0.2cm]
{\bf Key Point:} Each user $U_k$ can retrieve $\mathcal{S}_{k}$ from its connected cache-nodes and  can not get any information about $ \mathcal{S}_{j}$ where $j\in [K] \setminus \{k\}$.\\[0.2cm]
Since $S_{j,\mathcal{Q}}$ is locked by a key with the same length, it can be  seen from~\cite{CES} that  user $U_k$ cannot get any information about $S_{j,\mathcal{Q}} $.
 As a result, we also have
\begin{align}
%I(\mathcal{S}_{1},\ldots,  \mathcal{S}_{k-1}, \mathcal{S}_{k+1}, \ldots \mathcal{S}_{K} ;\mathcal{Z}_{U_k} ) =0.
I({\bf p}_{1}, \ldots,  {\bf p}_{K} ;\mathcal{Z}_{U_k}| {\bf p}_{k}, W_{1},\ldots, W_{N} )=0, \ \forall k \in [K].
\label{eq:private placement}
\end{align}

We  then illustrate the placement strategy by an example.
\begin{example}\rm
\label{exam-placement}
Consider the $(K,L,M,N)=(3,2,5/3,3)$ private multiaccess coded caching problem.
We design the following placement phase based on the $(K,L,M',N)=(3,2,1,3)$  non-private multiaccess coded caching scheme in \cite{RK}.

As   the scheme in \cite{RK}, we
divide each file $W_n$ where $n\in [3]$ into $3$ equal-length subfiles,
$
W_{n}=\{ W_{n,\{1\}},W_{n,\{2\}},W_{n,\{3\}}\}
$, where each subfile has $\frac{B}{3}$ bits.

For the first part of cached contents, for each $n\in [3]$,  the server places the subfiles $W_{n,\{1\}}$   into the cache-node $C_{1}$; places the subfiles $W_{n,\{2\}}$   into the cache-node $C_{2}$;   places the subfiles $W_{n,\{3\}}$  into the cache-node $C_{3}$. It can be checked that the memory size used for the first part of cached contents is equal to $M'=3\times \frac{1}{3}  =1$. Recall that in this example, user $U_1$ is connected to cache-nodes $C_1$ and $C_2$; user $U_2$ is connected to cache-nodes $C_2$ and $C_3$; user $U_3$ is connected to cache-nodes $C_3$ and $C_1$. Hence, we have $\mathcal{T}_{1}=\{ \{ 3\}\}$, $\mathcal{T}_{2}=\{\{1 \} \}$, $\mathcal{T}_{3}=\{\{ 2\} \}$. For the second part of cached contents,
      \begin{itemize}
      \item {\it Step 1.}  The server   generates $K=3$ vectors where for any $k\in [3]$, $\mathbf{p}_{k}=(p_{k,1},p_{k,2},p_{k,3}) \in \mathbb{F}_{2}^3$. By \eqref{eq-linear}, we have
\begin{gather*}
  S_{1,\{3\}}=p_{1,1} W_{1,\{3\}} \oplus p_{1,2} W_{2,\{3\}}\oplus p_{1,3}W_{3,\{3\}} , \\
  S_{2,\{1\}}= p_{2,1}W_{1,\{1\}}\oplus p_{2,2}W_{2,\{1\}}\oplus p_{2,3}W_{3,\{1\}} , \\
  S_{3,\{2\}}= p_{3,1}W_{1,\{2\}}\oplus  p_{3,2}W_{2,\{2\}}\oplus p_{3,3}W_{3,\{2\}} .
\end{gather*}
      \item {\it Step 2.}
      Let $A_{1,\{3\}},A_{2,\{1\}},A_{3,\{2\}}$ be random variables, each     of which is uniformly i.i.d. over   $\mathbb{F}^{  B/3 }_{2}$.
      Then cache-nodes $C_{k}$, $k\in [K]$ caches the following key sets respectively, %\footnote{\label{foot:field} We would like to perform the operations ``+'' and ``-'' on some finite field of sufficiently large size. In other words, with an abuse of notation, we use $W_J$ to denote both the binary multicast messages as well as its representation on some higher field size. In the proposed scheme, we need a field size with characteristic larger than $2$.}
\begin{gather*}
\mathcal{P}_1=\{S_{1,\{3\}}\oplus A_{1,\{3\}}, A_{3,\{2\}} \},\\
\mathcal{P}_2=\{S_{2,\{1\}}\oplus A_{2,\{1\}},A_{1,\{3\}} \},\\
\mathcal{P}_3=\{S_{3,\{2\}}\oplus A_{3,\{2\}}, A_{2,\{1\}}\}.
\end{gather*}
      \end{itemize}
In total,  each cache-node $C_{k}$ where $k\in[3]$ caches $\mathcal{Z}_{C_k}$, where
$\mathcal{Z}_{C_1}=\{W_{n,\{1\}} | \forall n \in[3]\} \cup \mathcal{P}_1$, $\mathcal{Z}_{C_2}=\{W_{n,\{2\}} | \forall n \in[3]\} \cup \mathcal{P}_2$ and $\mathcal{Z}_{C_3}=\{W_{n,\{3\}} | \forall n \in[3]\} \cup \mathcal{P}_3$. 
Hence,  the size of all the cached contents for each cache-node equals the capacity $  \frac{5}{3} =M B$.  Then the users can retrieve the contents as follows,
\begin{gather*}
\mathcal{Z}_{U_1}=\{W_{n,\{1\}},W_{n,\{2\}} | \forall n \in[3]\} \cup \mathcal{P}_1 \cup \mathcal{P}_2,\\
\mathcal{Z}_{U_2}=\{W_{n,\{2\}},W_{n,\{3\}} | \forall n \in[3]\} \cup \mathcal{P}_2 \cup \mathcal{P}_3, \\
\mathcal{Z}_{U_3}=\{W_{n,\{3\}},W_{n,\{1\}} | \forall n \in[3]\} \cup \mathcal{P}_3 \cup \mathcal{P}_1.
\end{gather*}
 Among  $S_{1,\{3\}}, S_{2,\{1\}},   S_{3,\{2\}}$, user $U_1$ can only recover $S_{1,\{3\}}$; user $U_2$ can only recover $S_{2,\{1\}}$; user $U_3$ can only recover $S_{3,\{2\}}$.
  \hfill $\square$
\end{example}

\subsubsection{Delivery Phase}
\label{subsect-delivery}
 User $U_k$ requests the file $W_{d_k}$ where $d_{k} \in [N]$. Now for any $k\in [K]$, we treat the following linear combination
 \begin{subequations}
\begin{align}\label{eq-virtual-linear}
&W_{d_k}\oplus ( \underset{n\in [N]}{\oplus } p_{k, n} W_{n}) \nonumber \\
&=p_{k, 1} W_{1}\oplus \cdots \oplus(p_{k,d_k}\oplus  1) W_{d_k}\oplus \cdots \oplus p_{k,N}W_{N} \\
&=q_{k, 1} W_{1}\oplus \cdots  \oplus q_{k,N}W_{N},
\end{align}
\end{subequations}
as a virtual and  independent file $W'_{d_k}$,
where $q_{k, j}= p_{k,j}$ for $j\in [N] \setminus \{d_k \}$ and $q_{k,d_k} \neq p_{k,d_k}$.   We also let ${\bf q}_k=(q_{k, 1},\ldots, q_{k, N})$.
%We split this file into subfiles as   the placement phase in scheme~1.
To preserve the privacy of the demand of each user $U_{k}$ where $k\in [K]$, we assume that the actual request of user $U_{k}$ becomes the virtual file $W'_{d_k}$.
% By splitting $W'_{d_k}$ into subfiles as   the placement phase in scheme~1,
 Hence, user $U_{k}$ should recover
 %the following linear combination, which is treated as the subfile of independent file $W'_{d_k}$,
\begin{align}\label{eq-subfile-virtual}
W'_{d_k, \mathcal{Q}} := q_{k, 1} W_{1,\mathcal{Q}}\oplus   \cdots \oplus q_{k,N} W_{N,\mathcal{Q}}, \  \forall \mathcal{Q} \in \mathcal{T}_{k}.
\end{align}

In the delivery phase, for the successful decoding, the server first broadcasts  ${\bf q}_1,\ldots, {\bf q}_K$ to the users;  thus, from the viewpoint of the other users  the demand of user $U_k$ is $W'_{d_k}$.
We  then use the delivery strategy of Scheme 1
to broadcast  messages, such that each user can recover $W'_{d_k, \mathcal{Q}}$.

Finally, user $U_k$ can decode the subfile $W_{d_k,\mathcal{Q}}$ where $\mathcal{Q} \in \mathcal{T}_{k}$ from  $W'_{d_{k},\mathcal{Q}}$ and $S_{k,\mathcal{Q}}$, thus it can recover its requested file $W_{d_k}$. In terms of the load incurred by the coefficients ${\bf q}_1, \ldots, {\bf q}_K$, it is easy to argue that in the information theoretic limit of large file size ($B \to \infty$) and fixed system parameters $K$ and $N$, the excess load of communicating ${\bf q}_1, \ldots, {\bf q}_K$ is vanishing, since the coefficients require $KN$ bits, and $KN/B$ can be made arbitrarily small for sufficiently large $B$. For the privacy, intuitively, from the viewpoint of the other users  the demand of user $U_k$ is $W'_{d_k}$.  The  information-theoretical proof for the privacy constraint in~\eqref{eq-privacy} is given in Appendix~\ref{sec:private placement}.
\begin{example}\rm
\label{exam-delivery}
 Let us return to Example \ref{exam-placement} with $K=N=3$ and $L=2$.
Assuming that the demand vector $\mathbf{d}=(1,2,3)$.
From \eqref{eq-virtual-linear}, we have three virtual files as follows,
\begin{gather*}
  W'_{1}= (p_{1,1}\oplus  1) W_{1}\oplus p_{1,2}W_{2} \oplus p_{1,3}W_{3} \\
  W'_{2}= p_{2,1}W_{1}\oplus (p_{2,2}\oplus  1) W_{2} \oplus p_{2,3}W_{3} \\
  W'_{3}= p_{3,1}W_{1}\oplus p_{3,2}W_{2} \oplus (p_{3,3}\oplus  1) W_{3}.
\end{gather*}
From \eqref{eq-subfile-virtual}, user $U_1$ requires the subfile $W'_{1,\{3\}}$ as follows,
$$
  W'_{1,\{3\}}= (p_{1,1}\oplus  1) W_{1,\{3\}}\oplus p_{1,2}W_{2,\{3\}} \oplus p_{1,3}W_{3,\{3\}};
$$
User $U_2$ requires the subfile $W'_{2,\{1\}}$ as follows,
\begin{gather*}
  W'_{2,\{1\}}= p_{2,1}W_{1,\{1\}}\oplus (p_{2,2}\oplus  1) W_{2,\{1\}} \oplus p_{2,3}W_{3,\{1\}};
\end{gather*}
User $U_3$ requires the subfile $W'_{3,\{2\}}$ as follows,
\begin{gather*}
  W'_{3,\{2\}}= p_{3,1}W_{1,\{2\}}\oplus p_{3,2}W_{2,\{2\}} \oplus (p_{3,3}\oplus  1) W_{3,\{2\}}.
\end{gather*}
According to the delivery strategy of non-private multiaccess coded caching scheme in \cite{RK}, server sends the following signal to users,
\begin{align}
  W'_{1,\{3\}}\bigoplus W'_{2,\{1\}}\bigoplus W'_{3,\{2\}}. \label{eq:example packet}
\end{align}
From~\eqref{eq:example packet}, user $U_1$ can decode $W'_{1,\{3\}}$, user $U_2$ can decode $W'_{2,\{1\}}$ and user $U_3$ can decode $W'_{3,\{2\}}$, respectively. Hence user $U_1$ can obtain $W_{1,\{3\}}$ from $W'_{1,\{3\}}$ and $S_{1,\{3\}}$, user $U_2$ can obtain $W_{2,\{1\}}$ from $W'_{2,\{1\}}$ and $S_{2,\{1\}}$,   user $U_3$ can decode $W_{3,\{2\}}$ from $W'_{3,\{2\}}$ and $S_{3,\{2\}}$. Finally each user can recover its requested file.
\end{example}
\section{Extension for $L\geq K/2 +1$}
\label{sec:other region}
When $L> K/2 +1$, after some slight modification of the proposed scheme in Section~\ref{sec-proposed}, we can guarantee the privacy constraint in~\eqref{eq-privacy}.   Specifically, the server encrypts each key, say $S_{k,\mathcal{Q}}$, $k\in [K]$ and $\mathcal{Q}\in \mathcal{T}_k$, into $\omega\geq 2$ shares by   Shamir's secret sharing  in \cite{Sharmir}, say $S^{(h)}_{k,\mathcal{Q}}$, $h\in[\omega]$, and then places each share into some appropriate cache-node.
By   Shamir's secret sharing, $S_{k,\mathcal{Q}}$ could be re-constructed from $\omega$ shares, and we cannot get any information about  $S_{k,\mathcal{Q}}$ by  any $\omega-1$ shares nor less.

Let us focus on user $1$. Let $w$ be the minimum value satisfying  that there exists some set $\mathcal{L}_{1}\subseteq [L]$ with cardinality $w$,  where $\mathcal{L}_{1}\not\subseteq \text{Mod}(\{k ,k +1,\ldots,k +L-1\},K)$ for each $k\in [2:K]$. We assume that the $h^{\text{th}}$ element in $\mathcal{L}_{1}$ is $j_h$, for each $h\in [\omega]$.
For each  $\mathcal{Q}\in\mathcal{T}_1$, we place the share  $S^{(h)}_{1,\mathcal{Q}}$ into cache-node $C_{j_h}$, where $h\in [\omega]$.
Hence, user $U_1$ can retrieve   $S_{1,\mathcal{Q}}$  from the cache-nodes  indexed by  $\mathcal{L}_{1}$, while any other user cannot get    any information about  $S_{1,\mathcal{Q}}$.

For each  user $k\in [2:K]$, we let $\mathcal{L}_k= \{\text{Mod}(j_1+k-1,K), \text{Mod}(j_2+k-1,K),\ldots, \text{Mod}(j_h+k-1,K)    \}$ and place the shares by a similar way as described above.  Then it can be seen that the {\bf Key point}  in Section~\ref{sec-proposed} holds.
By the symmetry,
the memory size of each cache-node is $M=M'+\omega(1-\frac{LM'}{N})$. Using the same delivery strategy in Section \ref{sec-proposed}, a $(K,L,M,N)$ private multiaccess coded caching scheme  where $M=M'+\omega(1-\frac{LM'}{N})$ with the same load and sub-packetization level as the original $(K,L,M',N)$ non-private scheme satisfying the three requirements in Theorem \ref{th-one}.
Note that when $L\leq K/2 +1$, we have $\omega=2$, the above scheme is equivalent to the proposed scheme in Section~\ref{sec-proposed}.

\section{Conclusion}
\label{sec-conclusion}
This paper formulated the multiaccess caching system with private demands, where each user cannot get any information about the demands of other users. We  then
proposed an approach to guarantee the privacy  for any non-private multiaccess coded caching scheme  satisfying some constraints. On-going works include the consideration on the private multiaccess coded caching problem with more general topologies.

 %show that a private multiaccess coded caching scheme can be obtain from a non-private multiaccess coded caching scheme.

\appendix
\section{Proof of  the Privacy for the Proposed Scheme}
\label{sec:private placement}
\iffalse
\subsection{Proof of~\eqref{eq:private placement}}
\label{subsec-proof-side}
{\color{blue}
For any user $U_k$, we have
\begin{subequations}
\begin{align}
&I(\mathcal{S}_{1},\ldots,  \mathcal{S}_{k-1}, \mathcal{S}_{k+1}, \ldots \mathcal{S}_{K} ;\mathcal{Z}_{U_k} ) \nonumber\\ \label{eq-infor-1}
\overset{\eqref{eq-memory} \eqref{eq-user-retri} \eqref{eq-linear} \eqref{eq-caching-set}}{=}&I({\bf p}_{1},\ldots,{\bf p}_{k-1},{\bf p}_{k+1},\ldots, {\bf p}_{K}; \notag \\
&\mathcal{P}_{\text{Mod}(k,K)},\ldots,\mathcal{P}_{\text{Mod}(k+L-1,K)}) \\
%&=&I({\bf p}_{1},\ldots,{\bf p}_{k-1},{\bf p}_{k+1},\ldots, {\bf p}_{K};\mathcal{S}_{k}) \label{eq-infor-key}\\
\end{align}
%\eqref{eq-infor-key} holds because user $U_k$ decoded the linear combinations in  $\mathcal{S}_{k}$ from $\mathcal{P}_{\text{Mod}(k,K)},\ldots,\mathcal{P}_{\text{Mod}(k+L-1,K)}$, and random variables are independent of the library files. \eqref{eq-infor-zero} because holds the vectors  are independent of the library files and the linear combination $\mathcal{S}_{k}$ generated by the vector ${\bf p}_{k}$ and files.
\end{subequations}

}
\fi
%\subsection{Proof of the privacy in \eqref{eq-privacy}}
%\label{subsec-proof-privacy}
We will prove that the private multiaccess coded caching scheme satisfies the privacy constraint in \eqref{eq-privacy}. Note that from~\eqref{eq:private placement} and that the demands are independent of the placement phase, we have
\begin{align}
&I({\bf p}_{1},   \ldots, {\bf p}_{K} ;\mathcal{Z}_{U_k}| {\bf d}, {\bf p}_{k}, W_{1},\ldots, W_{N} )=0;
\label{eq:private placement further}\\
&\text{and }I({\bf p}_{1}, \ldots,   {\bf p}_{K} ;\mathcal{Z}_{U_k}| d_k, {\bf p}_{k}, W_{1},\ldots, W_{N} )=0.
\label{eq:private placement further2}
\end{align}
With~\eqref{eq:private placement further2} and the demands are independent of the placement phase, we further have
\begin{align}
 I({\bf q}_{1}, \ldots, {\bf q}_{k-1}, {\bf q}_{k+1}, \ldots, {\bf q}_{K} ;\mathcal{Z}_{U_k}| d_k, {\bf p}_{k}, W_{1},\ldots, W_{N} )=0.
\label{eq:private placement further3}
\end{align}

Focus on user $U_k$ where $k\in [K]$, we have
\begin{subequations}
\begin{align}
&I({\bf d}_{-k}; X, \mathcal{Z}_{U_k}| d_{k}) \nonumber\\
&\leq   I({\bf d}_{-k};X, {\bf q}_{1},\ldots, {\bf q}_{K}, \mathcal{Z}_{U_k},W_{1},\ldots,W_{N} | d_k)   \\
&= I({\bf d}_{-k}; {\bf q}_{1},\ldots, {\bf q}_{K},\mathcal{Z}_{U_k},W_{1},\ldots,W_{N}|  d_{k})\label{eq-mul-one}\\
&= I({\bf d}_{-k}; \mathcal{Z}_{U_k}|  d_{k}, {\bf q}_{1},\ldots, {\bf q}_{K}  ,W_{1},\ldots,W_{N})\label{eq-mul-2}\\
&= H(\mathcal{Z}_{U_k}|  d_{k}, {\bf q}_{1},\ldots, {\bf q}_{K}  ,W_{1},\ldots,W_{N})\nonumber\\& -H(\mathcal{Z}_{U_k}| {\bf d} , {\bf q}_{1},\ldots, {\bf q}_{K}  ,W_{1},\ldots,W_{N}) \label{eq-mul-3}\\
&= H(\mathcal{Z}_{U_k}|  d_{k}, {\bf p}_{k}, {\bf q}_{1},\ldots, {\bf q}_{K}  ,W_{1},\ldots,W_{N})\nonumber\\& -H(\mathcal{Z}_{U_k}| {\bf d} , {\bf p}_{1},\ldots, {\bf p}_{K}  ,W_{1},\ldots,W_{N}) \label{eq-mul-4}\\
&=H(\mathcal{Z}_{U_k}| {\bf p}_{k}, d_{k},  W_{1},\ldots,W_{N})- H(\mathcal{Z}_{U_k}|  {\bf d}, {\bf p}_{k}  ,W_{1},\ldots,W_{N} )\label{eq-mul-5}\\
&=H(\mathcal{Z}_{U_k}| {\bf p}_{k},   W_{1},\ldots,W_{N})- H(\mathcal{Z}_{U_k}|{\bf p}_{k},   W_{1},\ldots,W_{N})\label{eq-mul-6}\\
&=0,
\end{align}
\end{subequations}
where \eqref{eq-mul-one} follows from the fact that $X$ is a function of $({\bf q}_{1},\ldots, {\bf q}_{K},W_{1},\cdots,W_{N} )$,~\eqref{eq-mul-2} follows from the fact that $({\bf q}_{1},\ldots, {\bf q}_{K}  ,W_{1},\ldots,W_{N})$ are independent of the users' demands,~\eqref{eq-mul-5} follows from~\eqref{eq:private placement further} and~\eqref{eq:private placement further3},~\eqref{eq-mul-6} follows from the fact that the placement   is independent of the  demands.


\begin{thebibliography}{1}

\bibitem{MAN}
M. A. Maddah-Ali and U. Niesen, ``Fundamental limits of caching", {\it IEEE Trans. Inform. Theory}, vol. 60, no. 5, pp. 2856-2867, 2014.
\bibitem{JCM}
M. Ji, G. Caire, and A. F. Molisch, ``Fundamental limits of caching in wireless D2D networks", {\it IEEE Trans. Inform. Theory}, vol. 62, no. 2, pp. 849-869, 2016.
\bibitem{JAJC}
M. Ji, A. M. Tulino, J. Llorca, and G. Caire, ``Caching in combination networks", {\it 49th Asilomar Conf. on Sig., Sys. and Comp}, Nov. 2015.
\bibitem{WJPD}
K. Wan, M. Ji, P. Piantanida, and D. Tuninetti, ``Caching in combination networks: Novel multicast message generation and delivery by leveraging the network topology", {\it IEEE Intern. Conf. Commun (ICC 2018)}, May. 2018.
\bibitem{NUMS}
N. Karamchandani, U. Niesen, M. A. Maddah-Ali, and S. Diggavi, ``Hierarchical coded caching", {\it IEEE ISIT}, Honolulu, HI, pp. 2142-2146, Jun. 2014.
%\bibitem{ART}
%A. Sengupta, R. Tandon, and T. C. Clancy, \emph{Fundamental limits of caching with secure delivery}, IEEE Trans. on Information Forensics and Security, vol. 10, no. 2, pp. 355¨C370, 2015.
%\bibitem{VPNV}
%V. Ravindrakumar, P. Panda, N. Karamchandani, and V. M. Prabhakaran, \emph{Private coded caching}, IEEE Trans. on
%Information Forensics and Security, vol. 13, no. 3, pp. 685¨C694, 2018.
%\bibitem{AZAY}
%A. A, Zewail, A. Yener. \emph{Device-to-device secure coded caching}, IEEE Transactions on Information Forensics and Security 15 (2019): 1513-1524.

\bibitem{Engle2017privatecaching}
F. Engelmann and P. Elia, A Content-Delivery Protocol, ``Exploiting the Privacy Benefits of Coded Caching", {\it 2017 15th Intern. Symp. on Modeling and Optimization in Mobile, Ad Hoc, and Wireless Networks (WiOpt)}, May. 2017.

\bibitem{WG}
K. Wan, and G. Caire, ``On the coded caching with private demands", {\it IEEE Trans. Inform. Theory}, vol. 67, no. 1, pp. 358-372, Jan. 2021.

\bibitem{VPA}
V. R. Aravind, P. Sarvepalli, A. Thangaraj, ``Subpacketization in coded caching with demand privacy", arXiv: 1909.10471, Sep. 2019.
\bibitem{SJB}
S. Kamath, J. Ravi, B. K. Dey, ``Demand-private coded caching and the exact tradeoff for N = K = 2", 2020 National Conference on Communications (NCC) (pp. 1-6), Feb. 2020.
%\bibitem{WHMDG}
%K. Wan, H. Sun, M. Ji, D. Tuninetti, and G. Caire, \emph{Fundamental limits of device-to-device private caching with trusted server},   arXiv:1912.09985. 2019
\bibitem{YT}
Q. Yan and D. Tuninetti, ``Fundamental Limits of Caching for Demand Privacy against Colluding Users", {\it IEEE Journal on Selected Areas in Inform. Theory}, vol. 2, no. 1, pp. 192-207, Mar. 2021.


\bibitem{arxivfunctionretrieval}
K. Wan,   H. Sun, M. Ji, D. Tuninetti, and G. Caire, ``On optimal load-memory tradeoff of cache-aided scalar linear function retrieval", {\it IEEE Trans. Inform. Theory}, vol. 67, no. 6, pp. 4001-4018, June. 2021.



\bibitem{JHNS}
J. Hachem, N. Karamchandanic, and S. N. Diggavi, ``Coded caching for Multi-level Popularity and Access", {\it IEEE Trans. Inform. Theory}, vol. 63, no. 5, pp. 3108-3141, May. 2017.

\bibitem{KKKBS}
K. K. K. Namboodiri, B. S. Rajan, ``Multi-Access Coded Caching with Secure Delivery", https://arxiv.org/abs/2105.05611, May. 2021.

\bibitem{CLWZG}
M. Cheng, Q. Liang, K. Wan, M. Zhang, G. Caire, ``A Novel Transformation Approach of Shared-link Coded Caching Schemes for multiaccess  Networks", arXiv:2012.04483v1[cs.IT], Dec. 2020
%\bibitem{RK}
%K. S. Reddy, N. Karamchandani, Rate-memory tradeoff for multiaccess coded caching with uncoded placement,{\em IEEE Trans. Commun}, Vol.68, No.6, pp. 3261-3274, 2020.
\bibitem{YCTC}
Q. Yan, M. Cheng, X. Tang, and Q. Chen, ``On the placement delivery array design in centralized coded caching scheme", {\it IEEE Trans. Inform. Theory}, vol.~63, no.~9, pp. 5821-5833, 2017.
\bibitem{RK}
K. S. Reddy, N. Karamchandani, ``Rate-memory tradeoff for multi-access coded caching with uncoded placement", {\it IEEE Trans. Commun}, Vol.68, No.6, pp. 3261-3274, 2020.
\bibitem{SR}
S. Sasi, B. S. Rajan, ``An Improved multi-access Coded Caching with Uncoded Placement", arXiv:2009.05377, Sep. 2020.

 \bibitem{SPE}
 B. Serbetci, E. Parrinello and P. Elia, ``Multi-access coded caching: gains beyond cache-redundancy", {\it IEEE Information Theory Workshop (ITW)},  Visby, Sweden, 2019, pp. 1-5.

\bibitem{MARB}
A. A. Mahesh, B. S. Rajan, ``Coded Caching Scheme with Linear Sub-packetization and its Application to multi-access Coded Caching", arXiv:2009.10923, Sep. 2020.
%\bibitem{OG}
%E. Ozfatura and D. G¨¹nd¨¹z, \emph{Mobility-Aware Coded Storage and Delivery}, IEEE Trans. Commun, Vol.68, No.6, pp. 3275-3285, 2020.
%
%\bibitem{MKR}
%P. Muralidhar, D. Katyal, B. Rajan,\emph{Maddah-Ali-Niesen Scheme for multiaccess Coded Caching}, arxiv:2101.08723v1 Jan. 2021


\bibitem{Sharmir}
A. Shamir, ``How to share a secret", {\it Communications of the ACM}, vol. 22, no. 11, pp. 612-613, Nov. 1979.

\bibitem{CES}
C. E. Shannon, ``Communication theory of secrecy systems", {\it The Bell System Technical Journal}, vol. 28, no. 4, pp. 656-715, Oct. 1949.

\end{thebibliography}
\end{document}